\pdfoutput=1
\documentclass[a4paper,fleqn,usenatbib]{mnras}

\usepackage{newtxtext,newtxmath}

\usepackage[T1]{fontenc}
\usepackage{ae,aecompl}

\usepackage{graphicx}	% Including figure files
\usepackage{amsmath}	% Advanced maths commands
\usepackage{amssymb}	% Extra maths symbols
\usepackage{bm}		% Bold maths symbols, including upright Greek
\usepackage{hyperref}
\usepackage{subfigure}
\usepackage{romannum}

\usepackage{comment} % For commenting out large portions of text

\title[PAHs in high-$z$ galaxies]{Detecting PAHs in high-$z$ galaxies in proxy:\\ Modelling physical conditions in an extremely strong damped Lyman-$\alpha$ absorber towards QSO SDSS J1143+1420 at $z$ = 2.323}

\author[Shaw \& Ranjan]{
Gargi Shaw$^{1}$\thanks{E-mail:gargishaw@gmail.com} and A. Ranjan$^{2}$
\\
$^{1}$Department of Astronomy and Astrophysics, Tata Institute of Fundamental Research, 1 Homi Bhabha Road, Mumbai 400005, India\\
$^{2}$Korea Astronomy and Space Science Institute, 776, Daedeokdae-ro, Yuseong-gu, Daejeon, 34055, Korea\\
}

\date{}

\pubyear{2022}

\begin{document}
\label{firstpage}
\pagerange{\pageref{firstpage}--\pageref{lastpage}}
\maketitle

%This is a simple template for authors to write new MNRAS papers. The abstract should briefly describe the aims, methods, and main results of the paper. It should be a single paragraph not more than 250 words (200 words for Letters).No references should appear in the abstract.

\begin{abstract}
We explore indirect methods to detect Polycyclic Aromatic Hydrocarbons (PAHs) in gas-rich, absorption-selected galaxies at high redshift. We look at the optical VLT/X-shooter observations of an intervening, extremely strong damped Lyman-$\alpha$ absorber (or ESDLA, with log ($N$(H I)$\gtrsim$21.7)) towards QSO SDSS J1143$+$1420 at redshift, $z_{ESDLA}$ = 2.323. Literature studies have shown that this ESDLA contains signatures of dust and diffuse molecular hydrogen and it was specifically chosen for our study due to its close spatial proximity (impact parameter, $\rm \rho = 0.6\,\pm\,0.3$kpc) with its associated galaxy. There is no direct detection of PAHs emission in the limited observations of infrared(IR)-spectra along this sight-line. Hence, we use CLOUDY numerical simulation modelling to indirectly probe the presence of PAH in the ESDLA. We note that PAHs need to be included in the models to reproduce the observed column densities of warm H$_2$ and C I. Thus, we infer the presence of PAHs indirectly in our ESDLA, with an abundance of PAH/H = 10$^{-7.046}$. We also measure a low 2175 \AA\, bump strength (E$\rm _{bump}\,\sim$0.03 - 0.19 mag) relative to star-forming galaxies by modelling extinction of QSO spectra by dust at the absorber rest-frame. This is consistent with the low PAH abundance obtained indirectly using CLOUDY modelling. Our study highlights the usage of CLOUDY modelling to indirectly detect PAH in high-redshift gas-rich absorption-selected galaxies.

\end{abstract}


%%%%%%%%%%%%%%%%%%%%%%%%%%%%%%%%%%%%%%%%%%%%%%%%%%

%%%%%%%%%%%%%%%%% BODY OF PAPER %%%%%%%%%%%%%%%%%%

\section{Introduction}

Extremely strong damped Lyman-$\alpha$ absorbers (ESDLAs) are defined as systems with large neutral hydrogen column density, 
log ($N$(H I)$\gtrsim$21.7) \citep{Noterdaeme2015}. They are observed towards bright background sources such as quasars (QSO) or
$\rm \gamma$-ray bursts (GRBs). DLAs (log ($N$(H I)$\gtrsim$20.3)) are assumed to statistically probe the outskirts (halo or the circumgalactic medium) of their associated galaxies 
\citep[see][]{Pontzen2008, Yajima2012, Guimaraes2012, Altay+13, RahmatiandSchaye2014, Shaw2016, Balashev2018, Rawlins2018}.
High-redshift ($\rm z\gtrsim$2) \textit{intervening}\footnote{Intervening absorption systems are defined as the ones \textbf{not associated} 
either with the QSO or the QSO host galaxy, i.e. $\rm z_{QSO}!\approx\, z_{abs}$.} QSO-ESDLAs are metal-rich, 
show higher dust content and the presence of diffuse H$_2$  more frequently than DLAs and similar to 
\textit{associated} GRB-DLAs \citep[see e.g.][]{Prochaska2008, Guimaraes2012, Bolmer2019, Ranjan2020, Shaw2020}, where the background source (a $\rm \gamma$-ray burst) and the gas clouds probed in absorption are from within the associated galaxy. Additionally, intervening QSO-ESDLAs have also been shown in the literature to statistically probe gas from within the star-forming disk of their associated galaxies \citep[][]{Kulkarni2012, Noterdaeme2014, Balashev2017, Ranjan2020, Ranjan2022arXiv}. 
%Since GRBs probe gas from within the associated galaxy, 
The ESDLA incident rate in GRBs is much higher \citep[$\sim$66\%, see][]{selsing2019x} compared with intervening QSO-ESDLAs \citep[$\sim$0.001\%, see][]{Noterdaeme2014} as quasar sightlines do not necessarily pass through an intervening galaxy. Yet, in contrast with GRB-ESDLAs, the associated galaxies of intervening QSO-ESDLAs are not selected based on their instantaneous star-forming nature.
Hence, intervening QSO-ESDLAs provide a unique opportunity to study the interstellar medium (ISM) in general, and the associated dust content, Polycyclic Aromatic Hydrocarbons (PAHs), specifically, within gas-rich galaxies at high-redshift. 

Dust grains, PAHs are ubiquitous and have been detected across the Milky Way \citep{{Draine2003},{Cohen1999}} and in various extragalactic sources \citep{{Calzetti2000},{Khramtsova2013}}. Their contribution is important to the heating budget and the general chemistry of the interstellar medium (ISM) \citep{Tielens2008}. Similarly, the cosmic-ray ionization 
also plays an important role in the chemistry of the ISM by controlling the ion-molecular chemistry and providing a source of heating. The cosmic-ray ionization rate $\chi_{CR}$ (s$^{-1}$) varies across the Milky Way galaxy \citep{Donnell1974, Indriolo2012, Neufeld2017} and is anti-correlated with PAH abundance \citep{Shaw2021}.
Moreover, \citet{Draine2007} noticed that the fraction of the dust mass in the form of PAHs correlates with metallicity. Further, \citet{Cortzen2019} showed that the luminosity of 6.2$\rm \mu$m PAH line emission correlates with CO luminosity for normal star-forming galaxies (SFGs) and starbursts (SBs) at all redshifts. Hence, it is important to probe PAH abundance in star-forming luminous galaxies \citep[see e.g.][]{Shivaei2022} as well as absorption selected gas-rich galaxies such as the ones probed by intervening QSO-ESDLAs\footnote{Here and further in this paper, we refer to `intervening QSO-ESDLAs' as ESDLAs for brevity.}. To this end, we perform a detailed numerical simulation for the high-redshift ESDLA towards QSO SDSS J1143$+$1420 at $z$ = 2.323 using the spectral synthesis code-CLOUDY \citep{Ferland2017, Shaw2005} and derive the underlying physical conditions. The particular ESDLA has been selected due to its close spatial proximity to its associated galaxy (most probable impact parameter, $\rm \rho = 0.6\,\pm\,0.3$kpc). The proximity and significant dust content (relative to other ESDLAs, with A$\rm _v\,\sim$ 0.23, see \citet{Ranjan2020}) makes it an ideal target for our intended first study to probe PAHs in high-redshift gas-rich galaxies. We note that PAH is not directly detected in infrared (IR) emission at the rest-frame of the ESDLA galaxy along the line-of-sight. Although, we also note that only two WISE bands were used for this conclusion as no other relevant IR data is available for this ESDLA as of the writing of this paper. 

The details of observations of the ESDLA towards QSO SDSS~J\,1143$+$1420 are presented together with the analysis of observed data in Section~\ref{sec:obs}. The interpretation of the observed results along with the numerical modelling using CLOUDY is discussed in Section~\ref{sec:cal}. We further summarize our discussions in Section~\ref{sec:sum}. Column densities are always stated in log and in the units of [atoms cm$^{-2}$]. For interpreting cosmic distances, the standard $\Lambda$CDM flat cosmology is used in this paper with $\rm H_0 = 67.8~km\,s^{-1} Mpc^{-1}$, $\rm \Omega_{\Lambda} = 0.692$ and $\rm \Omega_{m} = 0.308 $ \citep{Planck2016}.   

\section{Observations and gas properties} \label{sec:obs}

\subsection{Optical observations}
The ESDLA towards QSO SDSS~J114347.21$+$142021.60 (hereafter, J\,1143$+$1420, z$\rm _{QSO}$ = 2.583, z$\rm _{ESDLA}$ = 2.323) was initially detected in SDSS DR11 database by \citet{Noterdaeme2014}. The QSO was further observed twice in service mode under good seeing conditions (typically 0.7-0.8$\arcsec$) between April 2015 and July 2016 as a part of the ESO program ID 095.A-0224(A) with the multiwavelength medium-resolution spectrograph X-shooter \citep{Vernet2011} mounted at the Cassegrain focus of the Very Large Telescope (VLT-UT2) at Paranal, Chile. The observations were performed under good sky conditions (seeing $\sim\,0.76,\, 0.93$ and airmass $\sim 1.15$) for all exposures (exposure time $\rm \sim\,25$ minute). 
The full details about the observations and data reduction are given in \citet{Ranjan2020}. In this section, we briefly report the findings of \citet{Ranjan2020} and \citet{Ranjan2022arXiv} that are relevant for this work.

From observations and further analysis (fitting multiple FeII and SiII lines), they achieved an average S/N $\rm \simeq$ 51, 34 and 40 kms$\rm ^{-1}$ at the UVB, VIS and NIR arms of X-shooter, respectively. The main goals of the mentioned ESO program were to obtain the properties of the gas (such as metallicity, dust and H$_2$\, content in absorption) and possible emission signatures from the associated galaxy. \\

\subsubsection{Observed properties of the ESDLA gas}
The ESDLA towards QSO J\,1143$+$1420 presents strong H I absorption at a redshift, z$\rm _{abs}$ = 2.323, with a total H I column density, logN(H I) = 21.64$\pm$0.06, classifying it as an ESDLA within 1-$\rm \sigma$ uncertainty.  Inferring from singly ionized species, we obtain the gas-phase metallicity, [Zn/H] = -0.80$\pm$0.06.  

The column density estimates and limits (upper in case of non detection and lower in case the absorption profiles are saturated) of all species relevant for this work is given in Table  ~\ref{tab:table 2}. 

Diffuse H$_2$\, gas is also detected as Lyman-Werner (LW) band absorption in association with the neutral gas along the line of sight
with a total H$_2$\, column density, logN(H$_2$)=18.3$\pm$0.1 obtained using the first two (J = 0, 1) rotational levels of H$_2$. In Table ~\ref{tab:table 2}, along with J = 0, 1 levels, we also show a tentative estimate of the column density of the J = 2 levels. Due to the medium spectral resolution, low H$_2$\, column density, and contamination (both from the ${\rm Ly}\alpha$\, forest and the strong metal lines), it is difficult to robustly estimate the column density of higher (J$>$1) rotational levels.

C I and Cl I are both good tracers of H$_2$ but they are also not detected in the ESDLA, as is expected for the low N(H$_2$). 
%This is expected as the H$_2$\, column density itself is quite low. 
Table.~\ref{tab:table 2} shows the conservative 3-$\rm \sigma$ upper limit estimate of the column density of C I, and Cl I\, derived by fixing the $b$-value and redshift from H$_2$\, and modelling the noise in the prospective positions of C I and Cl I transitions. 

\subsubsection{Dust continuum modelling}
Dust at the ESDLA rest-frame can be modelled by applying the standard extinction laws \citep[e.g. from][]{Fitzpatrick_Massa_2007} taking extinction curves from \citet{Gordon2003} and applying it to QSO models \citep[such as][]{Selsing2016}. \citet{Ranjan2020} modelled the dust extinction for ESDLA towards QSO J\,1143$+$1420 using this method. For their work, they kept a simplistic model where the dust was only modelled at the absorber rest frame (and not at the QSO rest-frame) and all the other extinction curves parameters were fixed as given in the models from \citet{Gordon2003}. \citet{Ranjan2020} modelled the dust content, A$\rm _V\,\sim$ 0.23 mag with the best fit on the average
Small Magellanic Cloud (SMC) extinction template. \citet{Fitzpatrick_Massa_2007} note that one of the parameters in their template is directly related to a Lorentzian like bump centered at 2175\AA. The amplitude of the bump is, E$\rm _{bump}$=E(B-V)$\rm \times$(c3/$\rm \gamma^2$) with the total area underneath the bump, A$\rm _{bump}$=E(B-V)$\rm \times$(($\rm \pi\, \times$ c3)/ (2 $\rm\times\, \gamma$)\footnote{c3 and $\rm \gamma$ are parameters related to the 2175 \AA\, bump. Please see \citet{Fitzpatrick_Massa_2007} for a detailed explanation.}. The 2175\AA\, bump strength is shown to be directly related to PAH emission in star-forming galaxies \citep[see e.g.][]{Shivaei2022}. Although, \citet{Gordon2003} show that the fixed SMC template does not have a significant 2175 \AA\, bump. Since we are interested in studying the PAH content in this ESDLA, we show here the result of dust modelling using all average extinction laws mentioned in \citet{Gordon2003} such as average Galactic (MW), average Small Magellanic Cloud (SMC), the average Large Magellanic Cloud (LMC), and the average LMC super-shell (LMC2). We followed a method similar to \citet{Ranjan2020} and in our first iteration (let's call it Case-1), we only varied the dust at the absorber rest frame and the parameter for re-scaling the QSO template to match the observed QSO spectrum. We show the results of the fit to different average extinction laws in Table.~\ref{tab:table4} as Case-1. In Case-1, the 2175 \AA\, bump will always be fixed to the amount of dust modelled and the extinction law used. Hence, we performed a second, more robust iteration of fitting with the listed average extinction laws, where we included the two variables - c3 and $\rm \gamma$ as free parameters for the fit (at the ESDLA rest-frame) along with the dust at QSO (modelled with the `SMC' template) and the absorber rest-frame. Since c3 and $\rm \gamma$ are directly related to the 2175 \AA\, bump strength, we hope to get a more robust estimate of the latter using this technique. We show all the results from our second iteration as Case-2 in Table.~\ref{tab:table4}. \\

\begin{table*}
\centering
\begin{tabular}{ccccc}
\hline
Extinction Law & Galactic & SMC Bar & LMC Supershell & LMC Average \\
\hline
\hline
Case-1 &  &  &  &  \\
A$_V$(ESDLA) & 0.4 & 0.24 & 0.44 & 0.47 \\
%A$_V$(QSO) & 0 & 0 & 0 & 0 \\
$\rm {A_{bump}}$ & 0.68$\rm \pm$0.01 & 0.05 & 0.39 & 0.63$\rm \pm$0.01 \\
$\rm {E_{bump}}$ & 0.47$\rm \pm$0.01 & 0.03 & 0.26 & 0.43 \\
Reduced $\rm \chi^{2}$ & 56.31 & 22.7 & 21.9 & 40.98 \\
\hline 
Case-2 &  &  &  &  \\
A$_V$(ESDLA) & 0.18$\rm \pm$0.04 & 0.21$\rm \pm$0.03 & 0.19$\rm \pm$0.07 & 0.22$\rm \pm$0.05 \\
A$_V$(QSO) & 0.15$\rm \pm$0.01 & 0.03$\rm \pm$0.03 & 0.13$\rm \pm$0.02 & 0.14$\rm \pm$0.01 \\
$\rm {A_{bump}}$ & 0.48$\rm \pm$0.10 & 0.6$\rm \pm$0.09 & 0.54$\rm \pm$0.20 & 0.5$\rm \pm$0.11 \\
$\rm {E_{bump}}$ & 0.15$\rm \pm$0.03 & 0.19$\rm \pm$0.03 & 0.17$\rm \pm$0.06 & 0.16 $\rm \pm$0.04 \\
Reduced $\rm \chi^{2}$ & 18.14 & 18.56 & 18.08 & 18.29 \\
\hline
\end{tabular}
\caption{Comparison of dust content in QSO rest-frame (A$_V$(QSO)) and dust content (A$_V$(ESDLA), 2175 \AA\, bump strength (A$\rm _{bump}$) and 2175 \AA\, bump amplitude (E$\rm _{bump}$) at the absorber (ESDLA) rest-frame between difference extinction curve fits to the spectra towards QSO SDSS~J1143$+$1420 performed using different extinction laws as listed. Case-1 represents a constrained fit with only A$_V$(ESDLA) and QSO model flux re-scaling as free parameters. Case-2 represents more extended fit including A$_V$(QSO), $\gamma$ and c3 parameters that are responsible for the 2175 \AA\, bump (see main text for more details). Last row in each case shows the best-fit reduced $\rm \chi^{2}$. All values (except reduced $\rm \chi^{2}$) are in mag.}
\label{tab:table4}
\end{table*}

\begin{figure*}
\includegraphics[scale=0.32]{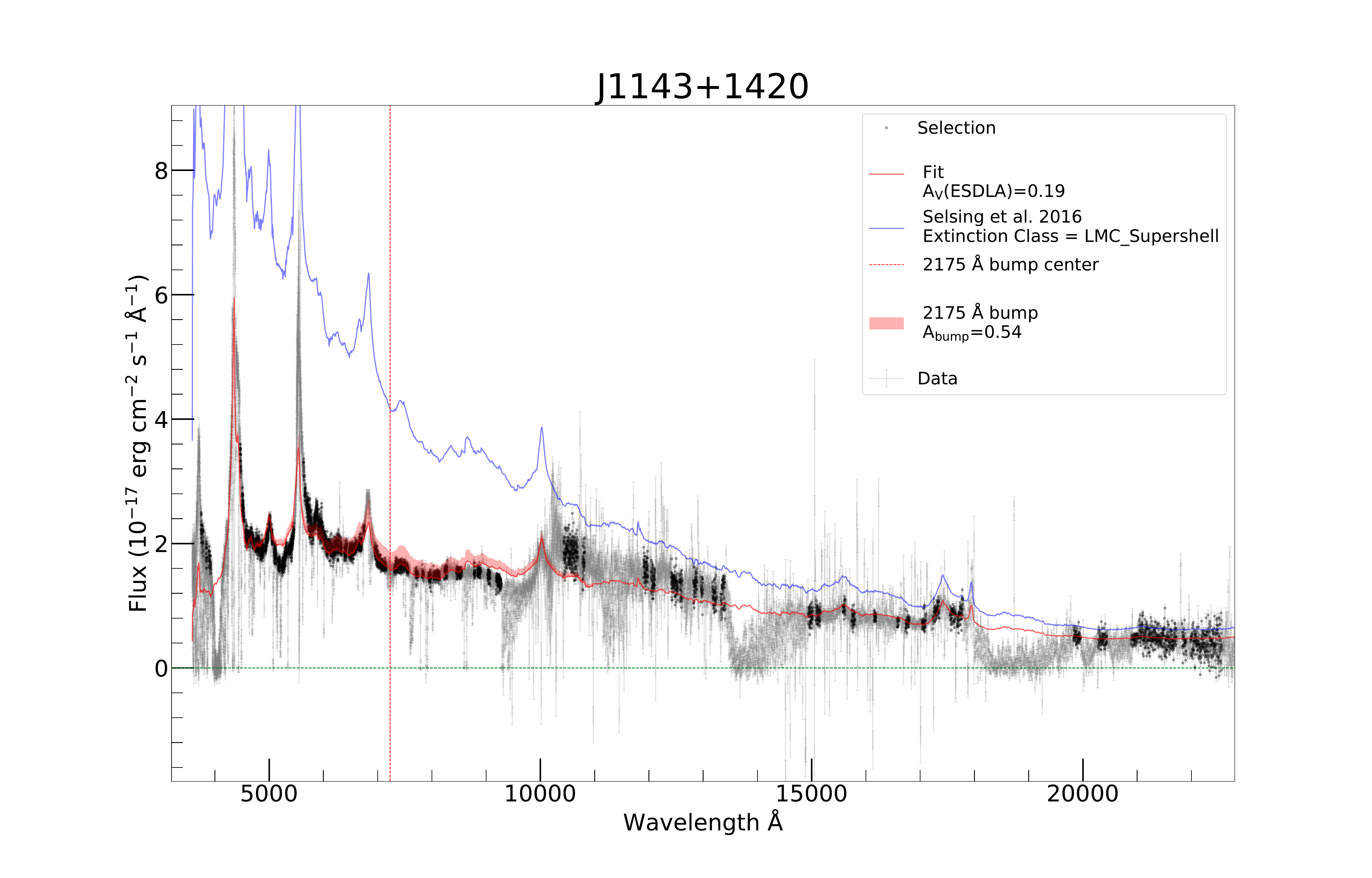}
\caption{Extinction curve fitting on the QSO X-shooter spectra towards SDSS~J1143$+$1420. The model QSO spectra is taken from \citet{Selsing2016}. The extinction at QSO rest-frame is also taking into account during the fit. The dust at ESDLA and QSO rest-frame has been modelled with the `LMC2' and `SMC' extinction types respectively. The extinction law templates were taken from \citet{Gordon2003} and implemented on a dust extinction modelling technique mentioned in \citet{Fitzpatrick_Massa_2007}. The impact of 2175\AA\, bump strength is shows as transparent red shaded region centered around 2175 \AA\, ESDLA rest-frame shown with a vertical red line.} 
\label{fig:esdla_dust}
\end{figure*}

We note that for acceptable fits (as checked by eye and having reduced $\rm \chi^{2}\,\lesssim$22), the 2175 \AA\, bump strength is higher in general for Case-2 fits. Figure~\ref{fig:esdla_dust} shows one of the Case-2 dust extinction fits towards the ESDLA using the LMC2 extinction curve and \citet{Selsing2016} QSO model. We note a slight mismatch in the figure between the fitted model and the QSO spectra around $\sim$10000\AA. We note that this effect is likely due to the change in the intrinsic shape of the QSO spectrum, $\Delta\,\beta$ \citep[see][and references there-in for more detail]{Krogager2016}. Changes in $\Delta\,\beta$ are degenerate with dust measurements. Since the effects of $\Delta\,\beta$ are not very pronounced for our case and the model matches the spectrum (within the 3-$\sigma$ error), we did not include this parameter as a variable in our fitting.

To conclude, since we cannot distinguish much between acceptable fits in Case-1 and Case-2, we take a conservative estimate for the 2175 \AA\, bump strength from all different listed values. Hence, we report the A$\rm _{bump}$ strength in this ESDLA system to be in the range, A$\rm _{bump}\sim$0.05 - 0.6 mag with the bump amplitude, $\rm {E_{bump}}\sim$0.03 - 0.19.

\subsubsection{Emission from associated galaxy}

[O III]$\lambda$5007 emission from the associated galaxy is detected in close spatial proximity (most probably impact parameter, $\rm \rho$ = 0.6$\pm$0.3 kpc, between the QSO sight-line and the centroid of emission) with a total line luminosity of 120($\pm$15)$\times$10$\rm ^{40}\,erg\,s^{-1}$ \citep[see][for more details]{Ranjan2020}. 
%Most DLAs are shown to probe the outskirts (halo or the circumgalactic medium) of their associated galaxy \citep[e.g.][]{Pontzen2008,RahmatiandSchaye2014, Yajima2012, Altay+13}. 
The close spatial proximity and less disruptive relative velocity ($\rm \Delta\, v\, \sim\,100\,kms^{-1}$) between absorption and emission indicate that the ESDLA cloud most likely lies within its associated galaxy. Hence, the ESDLA is ideal to perform a first case study on detecting PAHs in diffuse H$_2$\, clouds at high redshift.

\subsection{Infrared Observations}

We also looked for IR data (from WISE and Spitzer) for this ESDLA. Given the ESDLA rest-frame, only WISE bands - 3(12$\rm \mu$m), and 4(22$\rm \mu$m) with the angular resolution of 6.5 and 12 arcsec is available for our study. We report the non-detection of weak PAH emission \citep[rest frame, 3.1 to 3.7 $\rm \mu$m range and 6.0 to 6.9 $\rm \mu$m range, see][]{Salama2008} along the line of sight. Using WISE bands and looking at a spectra window of $\rm \sim\,500\,kms^{-1}$ at the ESDLA rest-frame, we report the upper limit in PAH emission luminosity along the mentioned bands to be, $<$0.05 and $<$1.83 (in units of 10$\rm ^{40}\, erg\, s^{-1}$) respectively.

\section{Calculations} \label{sec:cal}
All the simulations presented here are performed using the spectral synthesis code CLOUDY \citep{Ferland2017}. CLOUDY is a widely used microphysical code to predict and interpret astrophysical spectra over the entire electromagnetic range using a few input parameters. Further details about CLOUDY can be found in \citep{Ferland2013, Shaw2005, Shaw2020, Gay2012}.

\subsection{Modelling} \label{sub sec:model}
Our model is a time-independent model. We assume a plane-parallel gas cloud impinged by radiation from both sides \citep[similar to][]{{Ranjan2022arXiv},{Shaw2020},{Rawlins2018},{Shaw2016}}. 
Earlier \citet{Shaw2016} have experimented on modelling high-redshift DLAs by irradiating the cloud from both sides 
and from just one side. They have observed very minor differences between N(H I) and the H$_2$ level population predicted by 
these two cases. They also observed a small
difference of $\rm \sim$0.04 dex in the predicted N(CO) and C I fine structure population and finally used radiation impinging from both the sides. For consistency, we choose to use radiation impinging from both the sides. 
We irradiate the cloud from one side and stop our calculation at 
a depth in the cloud where the
total H I column density equals half of the observed H I 
column density. Finally, we multiply our predictions
by a factor of 2 to mimic the situation where the cloud is irradiated from both sides. The impinging radiation field consists of metagalactic radiation field \citep[see]{Khaire2019}, diffuse radiation similar to $\textit {in situ}$ star formation  and the Cosmic Microwave Background (CMB) at the appropriate redshift. Here, the diffuse radiation is denoted by $\chi$ (in terms of Habing field). The total hydrogen density, $n(H)$ cm$^{-3}$, includes all the forms of hydrogen-bearing species. In our model, $n(H)$, $\chi_{CR}$, $\chi$ are free parameters. 

Dust plays an important role in the ISM as it absorbs UV radiation and heats the gas through photoelectric heating. In a dusty environment, H$_2$ forms on dust grains and dust provides shielding from UV as well. We have included both graphite and silicate grains in the calculations with MRN \citep{Mathis1977} size distribution over ten size bins. According to MRN grain size distribution, the grain size follows a power-law distribution \begin{math} dn/da \propto a^{-3.5}, a_{min} \le a \le a_{max}\end{math}. Here `a' is the radius of the grain and `a$_{max}$' and `a$_{min}$' are 0.250 and 0.005 $\mu$m, respectively. It is assumed that Zn does not get depleted in interstellar dust grains but Fe does.
Hence, the metallicity is generally defined as, $[Zn/H]=log[N(Zn)/N(H)]\, -\, log[N(Zn)/N(H)]\odot$, while depletion is defined as, $[Fe/Zn] = log[N(Fe)/N(Zn)]\, -\, log[N(Fe)/N(Zn)]\odot$. Dust to gas ratio depends on metallicity through the relation, $10^{[Zn/H]}(1-10^{[Fe/Zn]}$). We fix the metallicity to -0.85, within the observed range of -0.8$\pm$0.06. We vary the dust to gas ratio taking into account the observed value of depletion, [Fe/Zn] = -0.54$\pm$0.03.  

%The observed value of depletion is, [Fe/Zn] = -0.54$\pm$0.03. We vary the dust to gas ratio within this observed range.

\citet{Tielens2008} determined the abundance of C locked up in PAHs containing 20--100 $C$ atoms as 14 parts per million $H$ atom. As the amount of PAHs is not exactly determined, we use it as a free parameter in our model. Here, we use size-resolved PAHs, distributed in 10 size bins. The size distribution of the PAHs is taken from \citet{Abel2008} with minimum and maximum radii of 0.00043 $\mu$m (30 $C$ atoms) and 0.0011 $\mu$m (500 $C$ atoms), respectively. For further details about our treatment of PAHs, see \citet{Shaw2021}. 

We perform a grid of models in the parameter space within an acceptable range. We then choose the parameter values that predict a better match to the observed values and finally fine-tune the relevant input parameters so that the observed data can be optimally modelled \citep[][]{Shaw2021}.

\subsection{Results}\label{sub sec:res} 

\begin{table}
	\centering
	\caption{Input  parameters for our best model}
	\label{tab:table 1}
	\begin{tabular}{lr} 
		\hline
	$\rm {Physical parameters}$ & $\rm {Values}$ \\ 
		\hline
		$\chi$ (G0) & 20\\
		$\chi_{CR}$ (s$^{-1}$)& 4$\times$ 10$^{-17}$ \\
		Density $n$($H$) ($cm^{-3}$) & 100\\
		Metals & 0.141 solar\\
		micro-turbulence (km s$^{-1}$) & 30\\
		Dust to gas ratio & 0.125\\
		PAH/H & 10$^{-7.046}$ \\
		$[Fe/H]$ & -1.35 \\
		$[Mn/H]$ & -1.49\\
		$[C/H]$ & -1.38 \\
		$[Zn/H]$ & -0.71 \\
		$[Ni/H]$ & -1.32 \\
		$[Ti/H]$ & -0.85\\
		$[Cr/H]$ & -1.2  \\
		$[Ar/H]$ & -1.74\\
		$[Cl/H]$ & -1.49\\
		$[Si/H]$ & -0.84\\
		$[O/H]$ & -0.85\\
		\hline
		\end{tabular}
        \end{table}

Table \ref{tab:table 1} lists the physical parameters of our best model. The metallicity and dust to gas ratio are 0.141 times solar and 0.125 times standard ISM value, respectively. 
The solar abundances are taken from \citep{Asplund2009}. We estimate $n$(H) = 100 cm$^{-3}$, radiation field $\chi$ = 20 (in the units of Habing field), cosmic-ray ionization rate $\chi_{CR}$ = 4$\times$10$^{-17}$ s$^{-1}$. 

Table \ref{tab:table 2} compares our predicted column densities with the observed ones. Our predicted column densities for all neutral gas species match within the observed range except O I. The O I column density for this system was derived by \citet{Ranjan2022arXiv} using only the O I$\lambda$1302 saturated profile. Inferring from the gas phase metallicity, the expected log($N$(O I) should be $\rm \sim$17.53$\pm$0.1. Hence, the modelled log($N$(O I)) = 17.45 is consistent (within 3-$\rm \sigma$ uncertainty) with the log($N$(O I) expected from the metallicity measurement. Hence, we conclude that the deviation from observations is rather a consequence of the O I$\lambda$1302 profile being possibly contaminated by the ${\rm Ly}\alpha$\, forest. We also reproduce H$_2$ column densities for the J = 0, 1, 2 levels.

Our model further predicts the presence of HeH$^+$ with an observable column density, $N$=10$^{11.56}$. Although, the HeH$^+$ lines emissions are outside the wavelength range of X-shooter and hence, cannot be directly detected.

\begin{table}
	\centering
	\caption{Comparison of predicted column densities (cm$^{-2}$) in log scale. Observations taken from \citet{Ranjan2020} and \citet{Ranjan2022arXiv}}
	\label{tab:table 2}
	\begin{tabular}{lll} 
		\hline
		$\rm {Chemical\, species}$ & $\rm {Observed}$ & $\rm {Predicted}$\\
	
		\hline
H I & 21.64$\pm$ 0.06 & 21.61\\
Zn II&13.47 $\pm$ 0.02& 13.46\\
Fe II & 15.78 $\pm$ 0.02 & 15.76\\
Cr II& 14.09 $\pm$ 0.02& 14.05\\
Si II & 16.30 $\pm$0.03 & 16.28\\
Ca II & $<$ 13.19 & 11.01 \\
Mn II & 13.59$\pm$0.01& 13.55\\
Ni II& 14.53 $\pm$0.02& 14.51\\
Ti II& $\approx$ 13.68 & 13.71\\
O I& $>$18 & 17.45 \\
O I$^{*}$& $<$19.16 & 17.45 \\
H$_2$ & 18.30$\pm$ 0.10 & 18.40 \\
H$_2$(0,0) & 17.77 $\pm$ 0.17 & 17.67\\
H$_2$(0,1) & 18.21 $\pm$ 0.07 & 18.26\\
H$_2$(0,2) & $\approx$17.24 & 17.25 \\
C I & $<$12.71& 12.67\\
Ar I & $<$15.2 & 14.27\\
Cl I &$<$12.68& 11.95\\		
		\hline
		\end{tabular}
        \end{table} 

\begin{figure}
\includegraphics[scale=0.5]{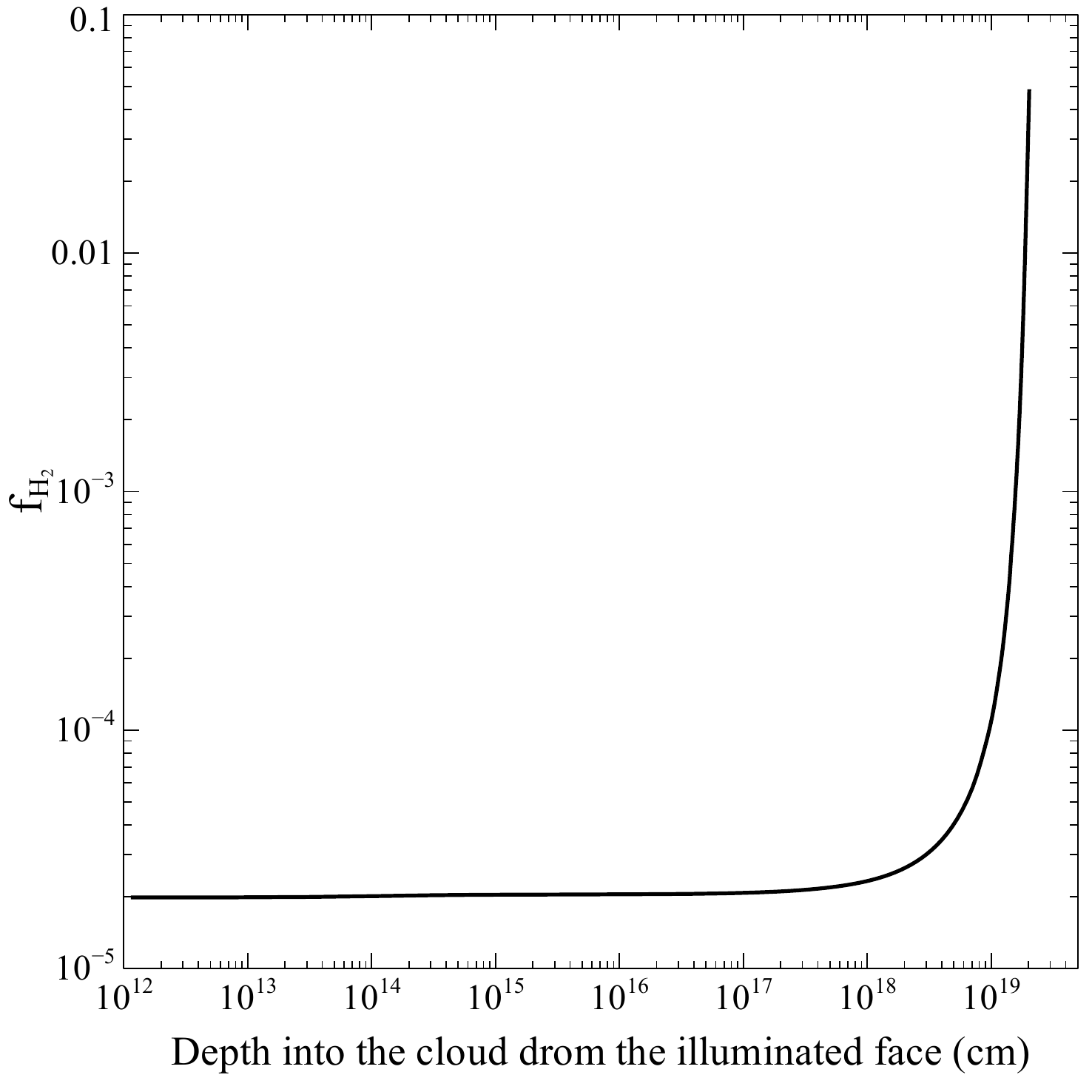}
\caption{Variation of the molecular fraction across the ESDLA. This represents one half of the cloud, the other half has a symmetrical profile to it.} 
\label{fig:structure_esdla}
\end{figure}

Fig.\ref{fig:structure_esdla} shows the variation of molecular fraction f$_{H_{2}}$ across the ESDLA. Here, f$_{H_{2}}$ is defined as,
\begin {equation}
f_{H_{2}} =\frac{2\times H_2}{2\times H_2 + H^0 +H^+}.
\end {equation}
Fig.\ref{fig:structure_esdla} depicts one half of the ESDLA as we stop our calculation at a depth 
in the cloud where the total H I column density equals half of the observed H I column density. 
The other half of the ESDLA has a symmetrical profile. It is clear from the plot that the cloud is mostly atomic as the molecular fraction is low, though it rises rapidly deep in the cloud.

\begin{figure}
\includegraphics[scale=0.5]{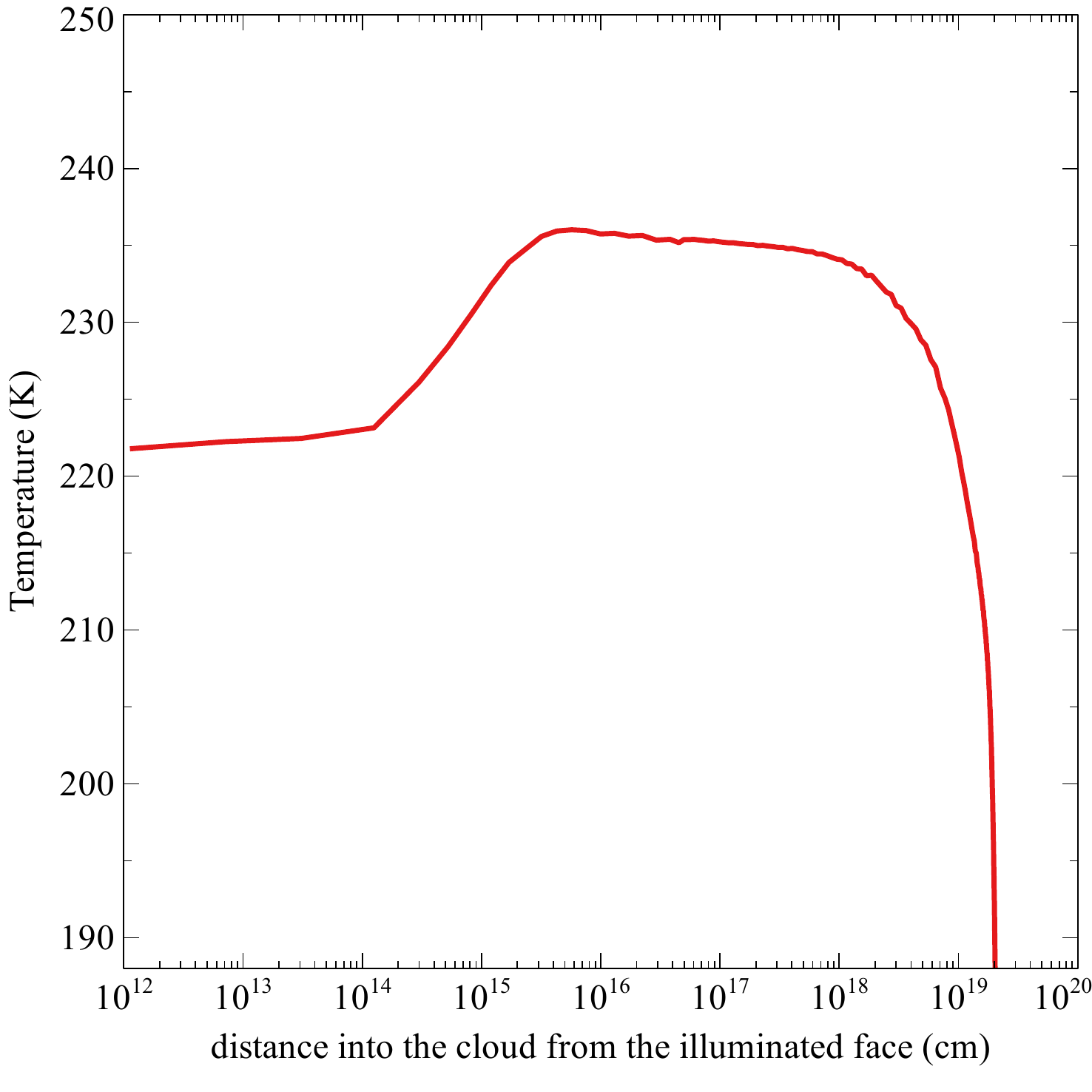}
\caption{Variation of the kinetic temperature across the ESDLA towards QSO SDSS J1143$+$1420. This represents one half of the cloud, the other half has a symmetrical profile to it.} 
\label{fig:esdla_pah}
\end{figure}

The J = 0 and 1 levels of H$_2$ are generally thermalised in a collisionally dominated region. 
The corresponding excitation temperature, T$_{10}$,  can be used to determine the
surrounding gas temperature in such conditions. This excitation temperature, T$_{10}$, is
calculated using the following equation,
\begin {equation}
T_{10} =\frac{-170.5}{ln \frac{N (J = 1)}{9 N(J = 0)}} K
\end {equation}
Here ${N(J = 1)}$ and ${N(J = 0)}$ are column densities in the J = 1 and 0
levels of H$_2$, respectively. The temperature, derived using H$_2$(0,0) and H$_2$(0,1), lies within 98-270 K. We predict the kinetic temperature averaged 
over the thickness of the system to be 220 $K$. Figure~\ref{fig:esdla_pah} shows the variation of the kinetic 
temperature across the ESDLA.  CLOUDY determines temperature from heating and cooling consisting of various 
physical processes. The kinetic temperature varies from 222 $K$ at the illuminated face to 188 $K$ at the centre of ESDLA. 
We find that the heating is dominated by photoelectric heating and the cooling is dominated by Ni II and C I 
recombination cooling. Figure~\ref{fig:esdla_pah} depicts one half of the ESDLA similar to Figure~\ref{fig:esdla_pah} with the other half of the ESDLA having a symmetrical profile. We predict the total extension of this ESDLA to 
be $\approx$ 13.55 pc assuming a plane-parallel geometry. 

As a test, we remove PAHs from our model while keeping all the other parameters the same. 
It creates a twofold effect. First, it affects the H$_2$ formation. H$_2$ forms on dust grains as well as on PAHs. Hence, there is a decrease in total surface area 
for H$_2$ formation. As a result, H$_2$ column density decreases to 10$^{15.56}$. Second, we notice 
that  the absence of PAHs makes the gas cooler due to less photoelectric heating. The kinetic temperature averaged over the thickness of the ESDLA comes down to 75.8 $K$. To increase the temperature we tried alternatively to increase $\chi_{CR}$ to produce more heating through cosmic rays. We need to increase $\chi_{CR}$ to 3$\rm \times\,10^{-15}\,s^{-1}$ to recover the temperature. However, the updated model is unable to reproduce the observed H$_2$ column density, and increases the C I, column density beyond its observed limit. We also tried to modify $n(H)$, $\chi$ and $\chi_{CR}$ among other parameters to reproduce the observed column densities in the absence of PAHs. We note that none of our PAH-absent models could reproduce all the observed column densities simultaneously.
%without considering PAHs in our model. 

\section{Discussion and Summary} \label{sec:sum}

We attempt to indirectly detect the presence of PAH in absorption selected gas-rich galaxy at high redshift. To this end, we look for multi-wavelength observations of an extremely strong damped Lyman-$\alpha$ absorber towards QSO SDSS J\,1143$+$1420 at $\rm z_{abs}$=2.323 which is at a very close spatial distance (most probable impact parameter, $\rm \rho = 0.6\,\pm\,0.3$kpc) relative to its associated galaxy. We perform a detailed numerical photo-ionization modelling of the observed gas cloud using the spectral synthesis code, CLOUDY. This source exhibits strong Ly$\alpha$ absorption, strong singly ionized metal lines and a diffuse H$_2$ absorption signature. We were able to reproduce all the observed column densities of all the observed neutral gas species. Although, to reproduce the column density of diffuse H$_2$ and C I (which traces diffuse H$_2$), we had to include PAH in our models. 
Literature studies show that intervening QSO-ESDLAs likely probe gas from within the star-forming disk of their associated galaxies. Hence, our work highlights a new method to indirectly estimate the presence and abundance of PAH in absorption-selected galaxies.  \\

Our main conclusions from this work are listed below:
\begin{itemize}
\item We estimate various underlying physical parameters for the ESDLA towards QSO SDSS J\,1143$+$1420 such as total hydrogen density $n$(H) = 100 cm$^{-3}$, diffuse radiation field $\chi$ = 20 (in the units of Habing field), cosmic-ray ionization rate 
$\chi_{CR}$ = 4$\times$10$^{-17}$ s$^{-1}$.
\item We predict the total thickness of this ESDLA to be $\approx$ 13.55 pc assuming plane-parallel geometry. 
\item The gas temperature averaged over the total length of the cloud is 220 $K$. 
\item We report the non-detection of weak PAH emission at ESDLA rest frame, 3.1 to 3.7 $\rm \mu$m range and 6.0 to 6.9 $\rm \mu$m range along the line of sight, with PAH emission luminosity along the mentioned bands to be, $<$0.05 and $<$1.83 (in units of 10$\rm ^{40}\, erg\, s^{-1}$) respectively. We note that there are no other relevant observations for direct PAH emission detection along the ESDLA line of sight as of the writing of this paper.

\item We note that our best model requires PAHs with number of PAHs per hydrogen, PAH/H = 10$^{-7.046}$. This provides indirect evidence of the presence of PAHs along this line of sight. 

\item Our model predicts HeH$^+$ with an observable column density of 10$^{11.56}$ cm$^{-2}$.

\item Literature studies \cite[see e.g.][]{Shivaei2022} show a correlation between the 2175\AA\, bump (A$\rm _{bump}$) and the PAH emission for star-forming galaxies. Our study shows low observed A$_{bump}$ strength (A$\rm _{bump}\,\sim$0.05 - 0.6 mag, E$\rm _{bump}\,\sim$0.03 - 0.19 mag) relative to star-forming galaxies (E$\rm _{bump}\,\sim$0.1 - 0.5 mag) which is consistent with our low modelled PAH abundance.
However, a follow-up sample study of PAH in ESDLA galaxies, similar to \citet{Shivaei2022}, would be required in identifying any quantitative relation between PAH abundance and A$_{bump}$ strength in gas-rich ESDLA environments. Additionally, a similar statistical study of associated GRB-ESDLAs will be helpful in probing PAH in gas-rich environments that have higher instantaneous star-formation rate \citep[see e.g.][]{Lyman2017} compared with intervening QSO-ESDLAs.  

\item The higher $J-$ levels of H$_2$ are sensitive to $\chi_{CR}$. However, this modelling is done with medium-resolution data where the high-$J$ ($J>$2) column densities are not reliable. We note that the study could further be advanced by observing the ESDLA using a high spectral resolution instrument such as VLT/UVES. We also note that a direct observation of PAH emission lines can be done in the future using IR space telescopes such as the James Webb Space Telescope (JWST). 

\end{itemize}

\section{Acknowledgements}

We thank the referee for their constructive comments and valuable suggestions on our paper. GS acknowledges WOS-A grant from the Department of Science and Technology (SR/WOS-A/PM-2/2021). 
AR acknowledges support from the National Research Foundation of Korea (NRF) grant funded by the Ministry of Science and ICT (NRF-2019R1C1C1010279).

\section{Data Availability}
Simulations in this paper made use of the code CLOUDY (c17.02), which can be downloaded freely at https://www.nublado.org/. 
The model generated data are available on request.

\bibliographystyle{mnras}
\bibliography{ESDLA_PAH} % if your bibtex file is called example.bib

%%%%%%%%%%%%%%%%%%%%%%%%%%%%%%%%%%%%%%%%%%%%%%%%%%

% Don't change these lines
\bsp	% typesetting comment
\label{lastpage}
\end{document}